\documentclass[english,structabstract]{aa}
\usepackage{mathptmx}
\usepackage[T1]{fontenc}
\usepackage[utf8]{inputenc}
\usepackage{babel}
\usepackage{graphicx}
\usepackage[unicode=true]
 {hyperref}
\usepackage{breakurl}

\makeatletter

\newcommand{\lyxdot}{.}

\bibpunct{(}{)}{;}{a}{}{,} % to follow the A&A style
\usepackage{hyperref}
\usepackage{longtable}
\usepackage{lscape}
\usepackage{txfonts}
\newcommand{\teff}{T_{\mathrm{eff}}}
\newcommand{\logg}{\log g}
\newcommand{\feh}{\left[\mathrm{Fe}/\mathrm{H}\right]}
\newcommand{\ltaur}{\log\tau_{\mathrm{Ross}}}
\newcommand{\taur}{\tau_{\mathrm{Ross}}}

\newcommand{\vzrms}{v_{z,\mathrm{rms}}}

\newcommand{\hav}{\left\langle \mathrm{3D}\right\rangle}

\newcommand{\angs}{\mathring{A}}
\newcommand{\eqw}{W_\lambda}
\newcommand{\lgf}{\log gf}

\newcommand{\fei}{\ion{Fe}{i}}
\newcommand{\feii}{\ion{Fe}{ii}}

\newcommand{\xex}{\chi_{\mathrm{exc}}}

\titlerunning{The Stagger-grid -- V. Fe line shapes, shifts and asymmetries}
\authorrunning{Z. Magic et al.}

\@ifundefined{showcaptionsetup}{}{%
 \PassOptionsToPackage{caption=false}{subfig}}
\usepackage{subfig}
\makeatother

\begin{document}

\title{The \textsc{Stagger}-grid: A Grid of 3D Stellar Atmosphere Models}

\subtitle{V. Fe line shapes, shifts and asymmetries%
\thanks{Full Table A.2 is available at the CDS via anonymous ftp to \protect\href{http://cdsarc.u-strasbg.fr}{cdsarc.u-strasbg.fr}
(\protect\href{http://130.79.128.5}{130.79.128.5}) or via \protect\href{http://cdsarc.u-strasbg.fr/viz-bin/qcat?J/A+A/???/A??}{http://cdsarc.u-strasbg.fr/viz-bin/qcat?J/A+A/???/A??},
as well as at \protect\href{http://www.stagger-stars.net}{www.stagger-stars.net}%
}}

\author{Z. Magic\inst{1,2}, R. Collet\inst{2} and M. Asplund\inst{2}}

\institute{Max-Planck-Institut für Astrophysik, Karl-Schwarzschild-Str. 1, 85741
Garching, Germany \\
\email{magic@mpa-garching.mpg.de} \and  Research School of Astronomy
\& Astrophysics, ANU Cotter Road, Weston ACT 2611, Australia}

\offprints{magic@mpa-garching.mpg.de}

\date{Received ...; Accepted...}

\abstract{}{We present a theoretical study of the effects and signatures of
realistic velocity field and atmospheric inhomogeneities associated
with convective motions at the surface of cool late-type stars on
the emergent profiles of iron spectral lines for a large range in
stellar parameters.}{ We compute 3D spectral line flux profiles under
the assumption of local thermodynamic equilibrium (LTE) by employing
state-of-the-art, time-dependent, 3D, radiative-hydrodynamical atmosphere
models from the \textsc{Stagger}-grid. A set of 35 real unblended,
optical $\fei$ and $\feii$ lines of varying excitation potential
are considered. Additionally, fictitious $\fei$ and $\feii$ lines
($5000\,\angs$ and $0,2,4\,\mathrm{eV}$) are used to construct general
curves of growth and enable comparison of line profiles with the same
line strength to illustrate systematical trends stemming from the
intrinsic structural differences among 3D model atmospheres with different
stellar parameters. Theoretical line shifts and bisectors are derived
to analyze the shapes, shifts, and asymmetries imprinted in the full
3D line profiles emerging self-consistently from the convective simulations
with velocity fields and atmospheric inhomogeneities.}{ We find systematic
variations in line strength, shift, width, and bisectors, that can
be related to the respective physical conditions at the height of
the line formation in the stellar atmospheric environment, in particular
the amplitude of the vertical velocity field.}{Line shifts and asymmetries
arise due to the presence of convective velocities and the granulation
pattern that are ubiquitously found in observed stellar spectra of
cool stars.}

\keywords{convection -- hydrodynamics -- line: formation -- line: profiles
-- radiative transfer -- stars: atmospheres -- stars: fundamental
parameters -- stars: late-type}

\maketitle

\section{Introduction\label{sec:Introduction}}

In recent years, capabilities for very high-resolution and very high
signal-to-noise spectroscopical observations have raised the level
of accuracy in stellar abundance analyses substantially \citep[e.g.,][]{Asplund:2005p7802,Melendez2009ApJ...704L..66M}.
In addition truly large-scale, comprehensive high-resolution spectroscopic
surveys are currently conducted, and further ones are planned. For
the accurate interpretation of these sterling data, improved theoretical
atmosphere models were also needed.

Cool stars are characterized by convective envelopes that extend to
the optical surface. The concomitant velocity field manifests itself
in the stellar photosphere with a typical granulation pattern that
imprints wavelength shifts and asymmetries in the observed spectral
line profiles \citep[e.g.,][]{Dravins:1981p15588,Nordlund:2009p4109}.
The strength of a spectral line depends mainly on the number of absorbers
(atomic level population), therefore, it is very sensitive to the
temperature due to exponential and power dependence of excitation
and ionization equilibria (in LTE: $\propto e^{-\chi/kT}$; see \citealt{Gray2005oasp.book.....G}
for more details on the theory of line formation in stellar atmospheres).
In addition, owing to the presence of convective velocities, additional
non-thermal broadening takes place in form of Doppler shifts. The
wings of spectral lines are formed in deeper layers close to the continuum
forming depth, while the cores are formed above in higher layers,
therefore, the line profile samples different heights with distinctive
physical conditions, in terms of velocity amplitudes, correlation
between temperature and density inhomogeneities, asymmetries between
regions with up- and downflowing material. This lead to characteristic
asymmetries in the emergent intensity and flux line profiles \citep[e.g.,][]{Asplund:2000p20875}.

Classical theoretical atmosphere models make use of several simplifications
in order to facilitate calculations with the computing power at hand
in the past \citep[e.g.,][]{Gustafsson:2008p3814,Cassisi:2004p1158}.
The treatment of convection is a partially challenging part in modeling
stellar atmospheres, since a complete theory of convection is still
absent. In one-dimensional (1D) modeling, simplified treatments of
convective energy transport have been adopted, such as the mixing-length
theory \citep[MLT; see][]{BohmVitense:1958p4822,Henyey:1965p15592}
or the full spectrum of turbulence model \citep{Canuto:1991p6553}
with a priori unknown free-parameters that has to be calibrated by
observations. Furthermore, for the 1D line formation calculations
the lack of knowledge on the convective velocity fields is partially
compensated by introducing two fudge parameters micro- and macroturbulence
to account for convective broadening of spectral lines \citep[e.g.,][]{Gray2005oasp.book.....G}.

For the precise modeling of realistic line profiles, including predicting
their asymmetries, one has to rely on realistic three-dimensional
(3D) atmosphere models, in which the convective velocity field emerges
from first principles, i.e. from the solution of the hydrodynamic
equations coupled with non-grey radiative transfer \citep[e.g.,][]{Nordlund:1982p6697,Steffen:1989p18861,Stein:1998p3801,Vogler:2003p11832,Nordlund:2009p4109,Freytag:2012p23073}.
A major application for 3D radiative hydrodynamic (RHD) atmosphere
models is the computation of synthetic full 3D line profiles or spectra
as post-processing based on the former in order to derive accurate
stellar parameters and abundances \citep{Asplund:2000p20875,Asplund:2000p20866,Asplund:2009p3308}.
The 3D RHD models demonstrated their predictive capabilities powerfully
in comparison with observed line profiles for several different stars.
\citet{Asplund:2000p20875} found almost perfect agreement between
observed solar iron line profiles and theoretical predictions without
any trends in the derived abundances with line strength. Furthermore,
comparisons of line shifts and asymmetries derived from high-resolution
spectroscopical observations of different types of cool stars advocated
additionally for the realistic nature of the theoretical 3D RHD models
\citep[e.g.,][]{Nordlund:1990p6720,Atroshchenko:1994p14010,AllendePrieto:2002p22280,Ramirez2008A&A...492..841R,Ramirez:2009p10905,Ramirez2010ApJ...725L.223R,Gray2009ApJ...697.1032G}.

With the present theoretical work we intend to tackle the following
key question: how do line properties vary with stellar parameters?
More specifically, we intend to analyze line shifts and asymmetries
carefully for a selection of $\fei$ and $\feii$ lines, in order
to better understand the variation of spectral line features with
stellar parameters. Iron lines are often considered most useful for
this purpose, since it is an abundant element with a very rich spectrum
of energy levels and transitions with a variety of strengths and accurate
atomic data at hand.

In Sec. \ref{sec:Methods} we explain the methods for the computations
of the 3D atmosphere models and line profiles. Subsequently, we discuss
the properties of lines in terms of shape, strength, width, and depth
(Sec. \ref{sec:Line-shape}, \ref{sub:Line-strength} and \ref{sub:Line-witdh-depth}),
as well as the line asymmetries and shifts (Sec. \ref{sec:Line-asymmetry}
and \ref{sec:Line-shift}). In Sec. \ref{sec:Height-of-line-formation},
we consider the physical conditions prevailing at the height of line
forming region and link them with the findings on the line profiles.
Finally in Sec. \ref{sec:Conclusions}, we summarize our results.

\section{3D atmosphere models and 3D spectral lines\label{sec:Methods}}

\begin{figure}
\includegraphics[width=88mm]{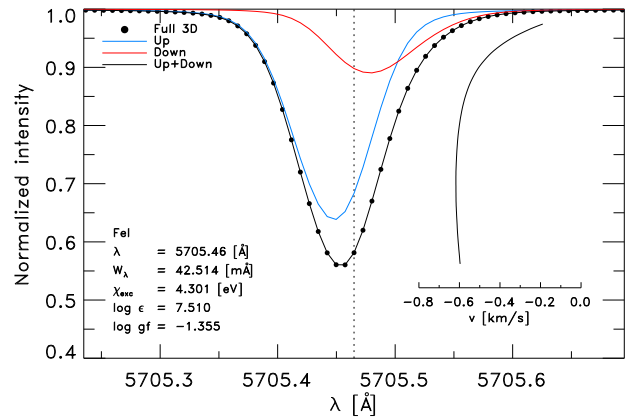}

\caption{Normalized disk-center ($\mu=1$) intensity profile as a function
of wavelength for a single $\fei$ line ($\lambda=5705.5,\,\xex=4.3$)
with intermediate strength synthesized using the solar simulation
(filled circles). The bisector information is plotted in the inset
on the right-hand side of the figure. The line is decomposed into
its contribution from upflows (blue) and downflows (red line), and
the sum of both (black solid line).}

\label{fig:line_up_down} 
\end{figure}
\begin{figure*}
\includegraphics[width=176mm]{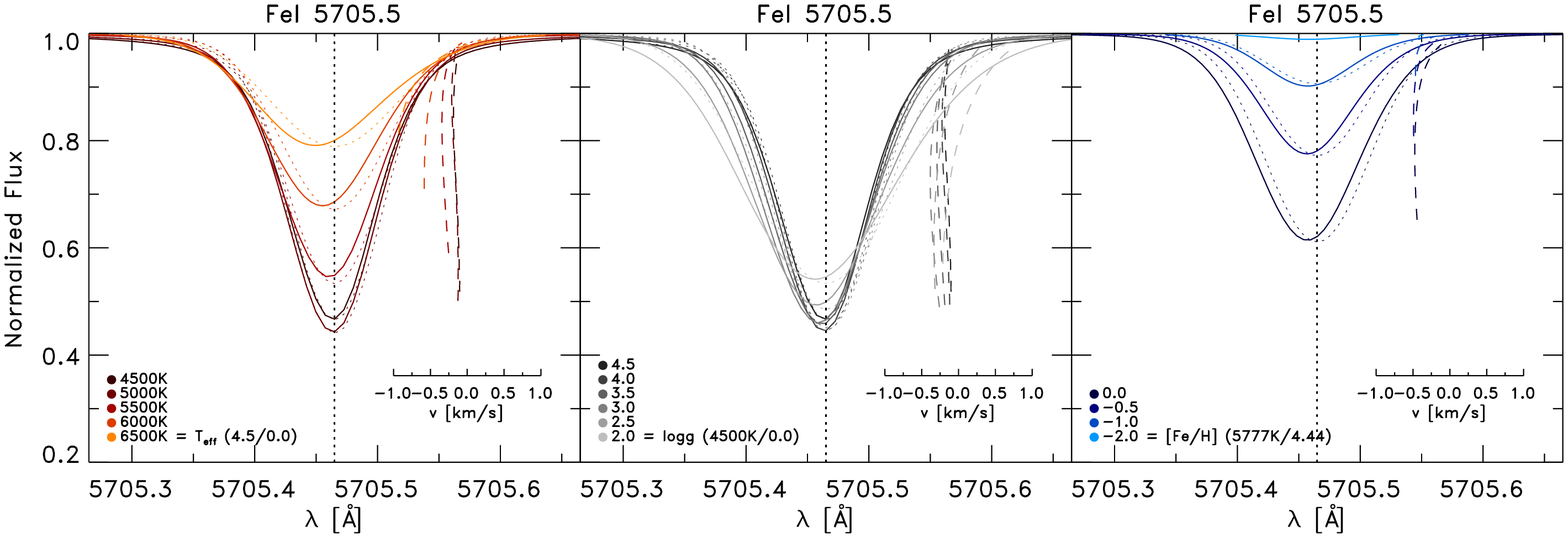}

\includegraphics[width=176mm]{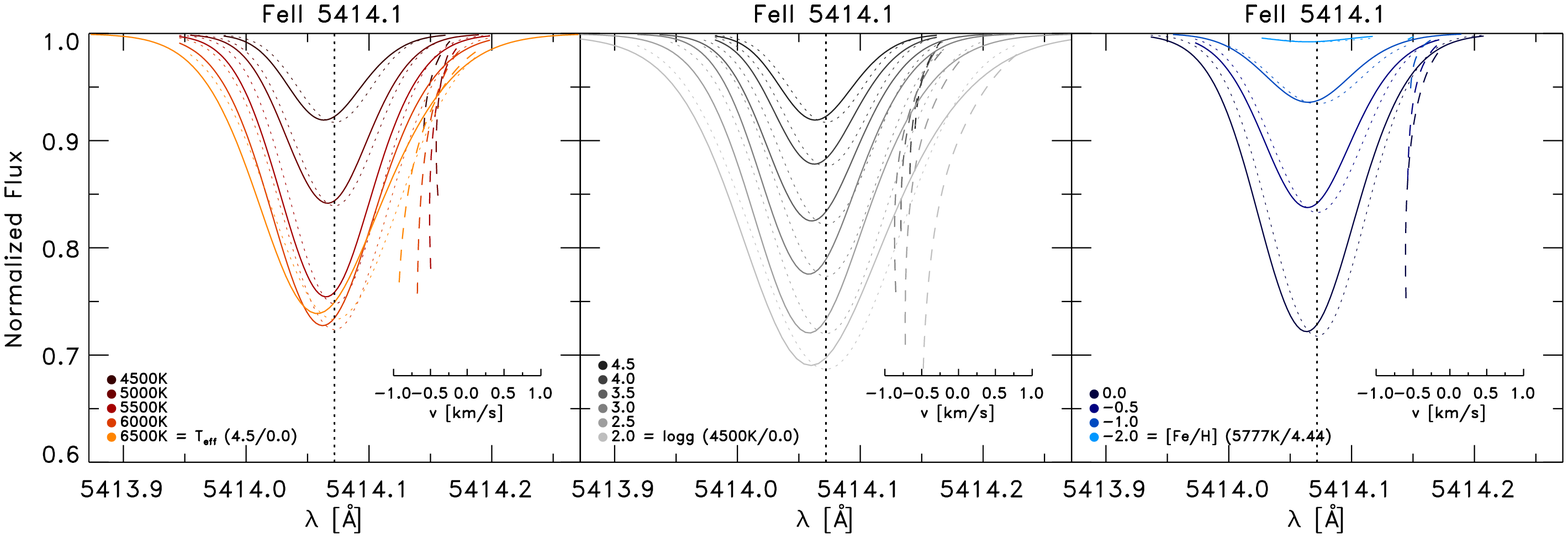}

\caption{Normalized flux vs. wavelength for a $\fei$ ($\lambda=5705.5,\,\xex=4.3$;
top panel) and $\feii$ line ($\lambda=5414.1,\,\xex=3.221$; bottom
panel) with solid lines for different stellar parameters with an enlarged
view of the bisector (dashed lines). In each column, only one stellar
parameter is varied, while the other two are fixed (indicated); \emph{left
panel}: effective temperature; \emph{middle panel}: surface gravity;
\emph{right panel}: metallicity. Furthermore, the respective lines
computed with the $\hav$ models are also shown (dotted lines). Note
the different the ordinates between the top and bottom rows.}

\label{fig:shape} 
\end{figure*}
We have computed the \textsc{Stagger}-grid, a large grid of 3D RHD
atmosphere models, which has been introduced in \citet[hereafter paper I]{Magic:2013}.
The stellar parameters of the \textsc{Stagger}-grid span in effective
temperature $\teff=[4000,7000]\,\mathrm{K}$ in steps of $500\,\mathrm{K}$,
surface gravity $\logg=\left[1.5,5.0\right]$ in steps of 0.5 dex,
and metallicity $\feh=\left[-4.0,+0.5\right]$ in steps of $0.5$
and $1.0\,\mathrm{dex}$%
\footnote{The metallicity is defined in respect to the solar value with $\feh=\log\left(N_{\mathrm{Fe}}/N_{\mathrm{H}}\right)_{\star}-\log\left(N_{\mathrm{Fe}}/N_{\mathrm{H}}\right)_{\odot}$.%
}. We employ a realistic equation-of-state (EOS) taken from \citep{Mihalas:1988p20892},
and the continuum and line opacity taken primarily from the MARCS
package \citep{Gustafsson:2008p3814}. In the \textquotedbl{}star-in-a-box\textquotedbl{}
setup, the rather small simulation box represents the statistical
properties of the stellar subsurface and atmosphere. The numerical
resolution of the Eulerian mesh of the box is $n_{xyz}=240^{3}$ for
all \textsc{Stagger}-grid models. The horizontal boundaries are periodic
and the vertical ones are open. The radiative transfer equation is
solved along long characteristics for 8 inclined rays plus the vertical
crossing for each grid-point at the surface of the simulation domain.
The effects of non-grey radiative transfer are accounted for in the
computations via the the opacity binning method \citep{Nordlund:1982p6697,Skartlien:2000p9857},
i.e. by sorting wavelengths into a number of opacity bins and properly
averaging the opacities in each bin before solving the radiative transfer
equation. We compute approximately two turn-over times, resulting
in 100 - 150 equidistant snapshots. Furthermore, we obtain the mean
$\hav$ stratifications from the latter, with the temporal and spatial
averaging methods as described in \citet[hereafter paper II]{magic:2013arXiv1307.3273M}.
For more details on the 3D atmosphere models we refer to paper I and
II.

For the calculations of the spectral absorption line profiles, we
utilize our 3D radiative transfer code \textsc{Scate} \citep[]{Hayek:2011p8560}.
We consider a subset of 20 equidistant snapshots from the complete
3D simulation sequence and solve the radiative transfer along four
polar $\mu$-angles, four azimutal $\varphi$-angles and the vertical
for $n_{\lambda}=101$ wavelength points by applying consistently
the same EOS and opacities as are used in the 3D atmosphere computations.
In order to ease the computational burden for the individual line
formation calculations, we assume local thermodynamic equilibrium
(LTE), i.e. $S_{\lambda}=B_{\lambda}$, and neglect scattering. Furthermore,
we reduce the number of columns in each direction from $240$ to 60,
we have sorted that this does not alter the line profile noticeably,
since the subsample is still large enough. The resulting intensity
profiles are spatially averaged for each snapshot. In order to compute
disk-averaged flux profiles, we integrate the various intensity profiles
at all inclined angles using a Gauss-Legendre quadrature scheme. Lastly,
we average the intensity and flux profiles temporally. In this work,
we prefer to discuss flux profiles over intensity profiles, since
the former are that what we observe in real stellar spectra except
for the Sun, whose surface can be spatially resolved.

For this investigation we employ $\fei$ and $\feii$ lines, which
are the same as considered by \citet{Asplund:2009p3308}. These lines
consist of carefully selected unblended lines. In Table \ref{tab:line_list}
we list the complete line list with their respective line parameters.
From the total of 35 lines, 26 are from neutral ($\fei$), while 9
lines are from singly ionized iron ($\feii$). As one can gather from
the line parameters, we cover the visible wavelength range with $\lambda=\left[4445.5,8293.5\right]\,\angs$,
and we match the ranges in lower excitation potential energy with
$\xex=\left[0.087,4.608\right]\,\mathrm{eV}$ and oscillator strength
with $\lgf=\left[-5.412,-1.355\right]$. For the solar simulation,
the line strengths range from $\eqw=\left[9.2,64.3\right]\,\mathrm{m\angs}$,
i.e. we cover from weak to intermediate line strength, however the
$\eqw$ varies with stellar parameters, and we get stronger lines
with a global maximum of $165.5\,\mathrm{m\angs}$.

In addition, we compute fictitious $\fei$ and $\feii$ lines for
a single wavelength ($\lambda=5000\,\angs$) covering three different
lower excitation potential energies ($\xex=0,2,4\,\mathrm{eV}$) in
order to facilitate a comparison between 3D models with different
stellar parameters based on the same line strengths. For the so-called
\textquotedbl{}curve-of-growth\textquotedbl{} method we consider ten
equidistant $\lgf$ values resulting in a range of line strengths
from weak to intermediate ($\eqw=5-100\,\mathrm{m\angs}$ for all
models). We then interpolate the computed line profiles to construct
a series of line profiles regularly distributed in line strength,
from $\eqw=10$ to $100\,\mathrm{m\angs}$ in steps of $10\,\mathrm{m\angs}$.
Finally, the line asymmetries (shift, width, depth and bisector) are
determined from the resulting interpolated line profiles. The line
shift is derived after the line profile is spline-interpolated to
a finer resolution around the line core, while the bisector is determined
for 100 points with spline interpolation.

\section{Line shape\label{sec:Line-shape}}

Real spectral absorption lines exhibit a more complex shape than just
a Gaussian or Lorentzian profile due to the properties of the convective
velocity field and inhomogeneities in the atmospheres of cool stars.
In order to elucidate the individual contribution from the granules
and the intergranular lane on the line shape and asymmetry, we show
in Fig. \ref{fig:line_up_down} a $\fei$ line computed from the solar
simulation considering a single snapshot. We separated the line profiles
of the individual columns into (bright) granules and (dark) intergranular
lane (up- and downflows) based on a threshold at 90\% from the mean
continuum intensity. As expected, upflows show stronger blue-shifted
profiles, while downflows have weaker, red-shifted lines \citep{Asplund:2000p20875}.
Furthermore, one can a notice distinct difference in the line depth
of the individual components, which unveils the fact that the effect
of the downflows is restricted mainly to the upper part of the bisector,
while the lower part is predominantly contributed by the upflows in
the granules.

In Fig. \ref{fig:shape}, we illustrate for a single $\fei$ and $\feii$
line (top and bottom panel respectively) overviews of profiles including
the bisectors, where one stellar parameter is varied, while the two
others are fixed, in order to illustrate the individual influence
of $\teff,\,\logg$ and $\feh$ on the line shape and asymmetry.

The variations of line profiles and asymmetries with stellar parameters
are systematic. Namely, the line strength decreases for $\fei$ (increasing
for $\feii$) with hotter $\teff$; the overall strength of spectral
lines decreases, naturally, with lower $\feh$; asymmetries become
more pronounced at higher $\teff$ and lower $\feh$ for both $\fei$
and $\feii$. The opposing trends of the $\fei$ and $\feii$ line
strength with $\teff$ stems from the ionization of neutral iron at
higher temperatures. In the case of the $\fei$ line, line strength
varies only marginally as $\logg$ decreases, while for $\feii$ line,
the line strength increases significantly. In fact, the $\feii$ line
strength shows always a clear rise for lower $\logg$, while for $\fei$
lines this is the case only for cooler $\teff$, and hotter ones show
even smaller trends with $\logg$. This behavior is easily understood
if one considers that when most of the iron is in singly ionized form
and $\fei$ is a minority species, the line opacity is proportional
to the electron number density which in turn is sensitive to the surface
gravity. Also, in the atmospheres of late-type stars, the main source
of continuum opacity in the optical is $\mathrm{H}^{-}$ and that
also scales proportionally to the electron number density. Since the
line strength of a normalized profile is proportional to the ratio
of line and continuum opacity, this means that, to first-order approximation
the dependence on electron density and, hence, on surface gravity
cancels out in the case of $\fei$ lines. Concerning the line shapes,
towards giants (lower $\logg$) the asymmetries increase considerably
for both lines due to the larger amplitude and velocity asymmetry
between up- and downflows (see Fig. \ref{fig:correlation_velocity_temperature}).
The wider and stronger line profiles in giants exhibit a more pronounced
redshift in the upper bisector (see lines with $\logg=2.0$ in middle
panel of Fig. \ref{fig:shape}) due to the increasing influence of
the contribution from downflows on the red wing. For the highest $\teff$
($6500\,\mathrm{K}$) or the lowest $\logg$ ($2.0$) the largest
span in asymmetry is achieved for both lines (see Fig. \ref{fig:shape};
also bottom panel of Fig. \ref{fig:mean_lineshift_max_bisector}).

The height of line formation is in general very important, namely
weaker lines show more pronounced line shifts and asymmetries, since
they tend to form in deeper layers, where the maximum velocities and
temperature contrasts happen ($\taur\sim1$; see Sec. \ref{sec:Height-of-line-formation}).
Stronger lines have their formation height shifted outwards where
the velocity and contrast is lower and less well anti-correlated (see
Fig. \ref{fig:correlation_velocity_temperature}). Similarly, $\feii$
lines are formed deeper than $\fei$, since their number density increases
in the deeper layers.

The wings are formed in deeper layers leading to a $/$-shape in the
bisector arising from the granules (see Fig. \ref{fig:line_up_down}),
while the line core originates from higher layers. Therefore, with
increasing line strength the line shift is receding after a maximum,
leading to the $\backslash$-shape to the typical C-shape of the bisector
(see Fig. \ref{fig:shape}). The line profile is often dominated by
the granules, since these exhibit a brighter intensity, steeper temperature
gradients and more importantly larger area contribution (filling factor)
compared to the downflowing intergranular lane. However, this correlation
decreases quickly above the optical surface, where convection ceases.
Furthermore, the line shape and bisector depend on the vertical velocity
field, its amplitude, asymmetry and the extent of overshooting into
convective stable layers. The radial p-mode oscillations generally
broaden the line profile, without however altering the overall line
strength noticeably, since the nearly Lagrangian vertical oscillations
do not influence the atmospheric stratification on an optical depth
scale.

Additionally, we included in Fig. \ref{fig:shape} also the symmetric
line profiles resulting from the corresponding mean $\hav$ models,
in order to depict the influence from the inhomogeneities and in particular
the vertical velocity field resulting from convection and granulation.
The homogenous $\hav$ models include micro- and macroturbulence in
order to yield the same line strength and depth respectively as the
considered full 3D line, therefore, one can isolate visually the Doppler
shifts arising from the realistic 3D velocity field. At cooler $\teff$
the line shape is more symmetric, and the effects of velocities are
rather small, while towards higher $\teff$ the differences grow more
apparent. Due to the Doppler shifts the 3D lines are more blue-shifted.

\section{Line strength\label{sub:Line-strength}}

\begin{figure}
\includegraphics[width=88mm]{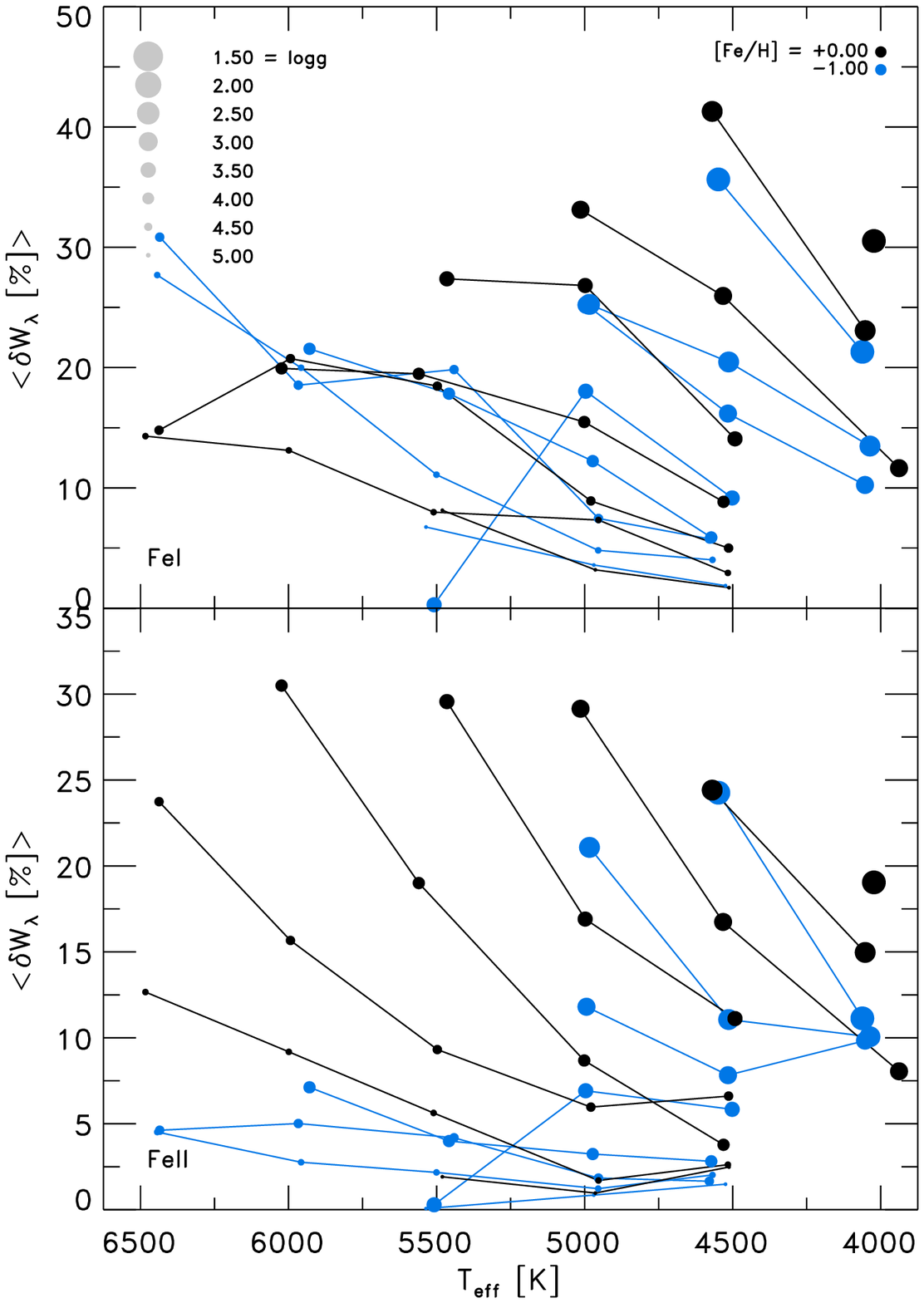}

\caption{\label{fig:diff_eqw}The relative difference between full 3D and $\hav$
line strength, $\delta\eqw$, vs. $\teff$ averaged over all $\fei$
and $\feii$ lines separately (top and bottom panel respectively).}
\end{figure}
We evaluate the relative difference $\delta\eqw=W_{\mathrm{3D}}/W_{\hav}-1$
between the predicted line strengths from the calculations with full
3D and with $\hav$ models without any microturbulence. In Fig. \ref{fig:diff_eqw},
we display the average difference separated in $\fei$ and $\feii$
lines with stellar parameters. We remark that the individual lines
exhibit distinctive values between different lines, however, the average
values depict qualitatively an overview of the variations. The difference
$\delta\eqw$ quantifies the effects of atmospheric inhomogeneities
as well as non-thermal Doppler broadening, since the $\hav$ models
include no velocity field or microturbulence. As expected the Doppler
broadening due to the convective velocities is enhancing the line
strength of the full 3D lines, with the consequence of the latter
being stronger than the $\hav$ lines. The enhancement in $\eqw$
is increasing for hotter $\teff$, lower $\logg$, and higher $\feh$,
which corresponds to the variation of the vertical rms-velocity (see
Fig. \ref{fig:psg_velocity_temperature}). The trends of $\delta\eqw$
with stellar parameters are between $\fei$ and $\feii$ in general
qualitatively rather similar. Lines with higher excitation potential
energy feature a smaller range in $\delta\eqw$.

\section{Line width and depth\label{sub:Line-witdh-depth}}

\begin{figure*}
\subfloat[\label{fig:width_depth}]{\includegraphics[width=88mm]{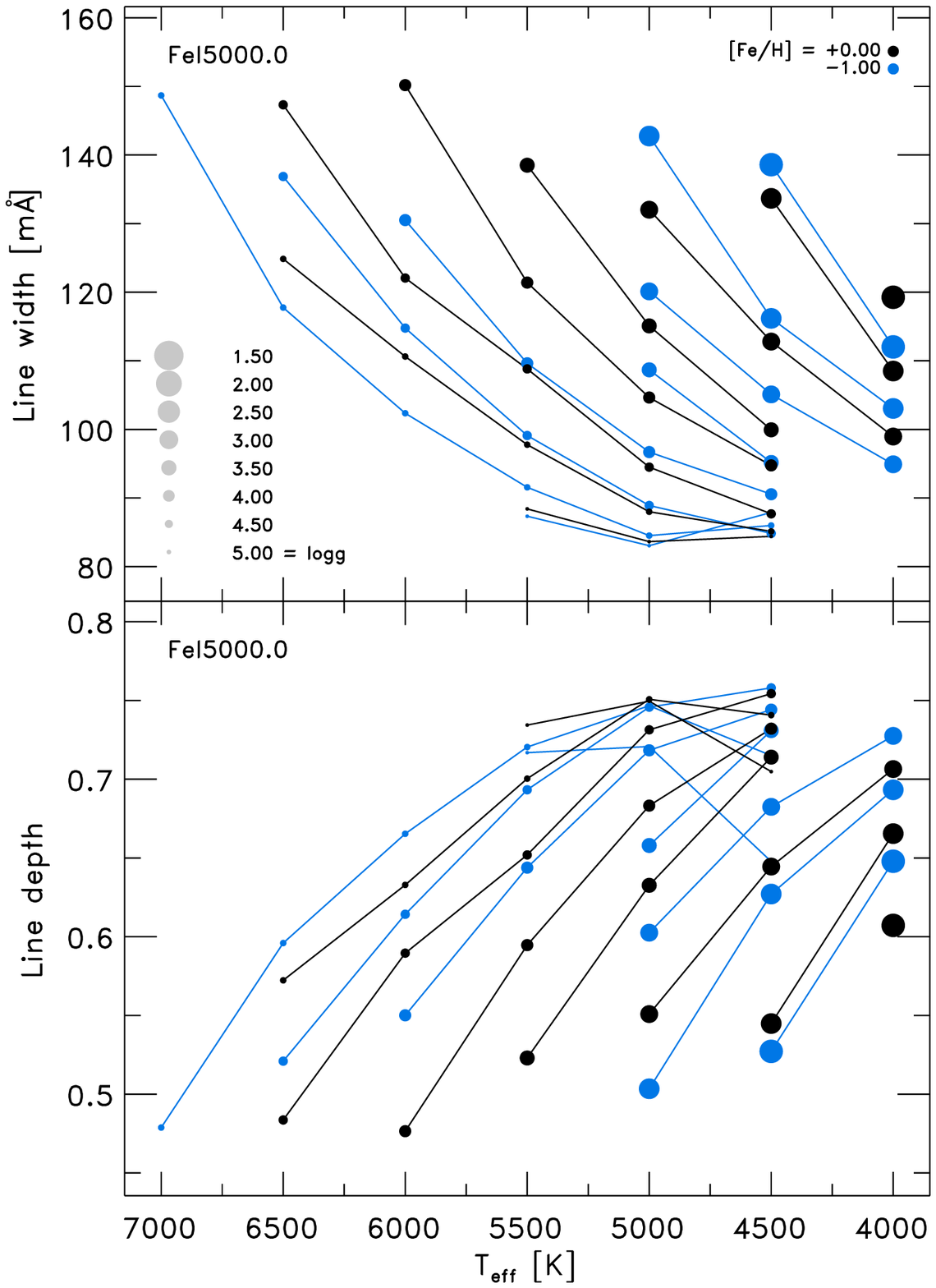}

}\subfloat[\label{fig:mean_lineshift_max_bisector}]{\includegraphics[width=88mm]{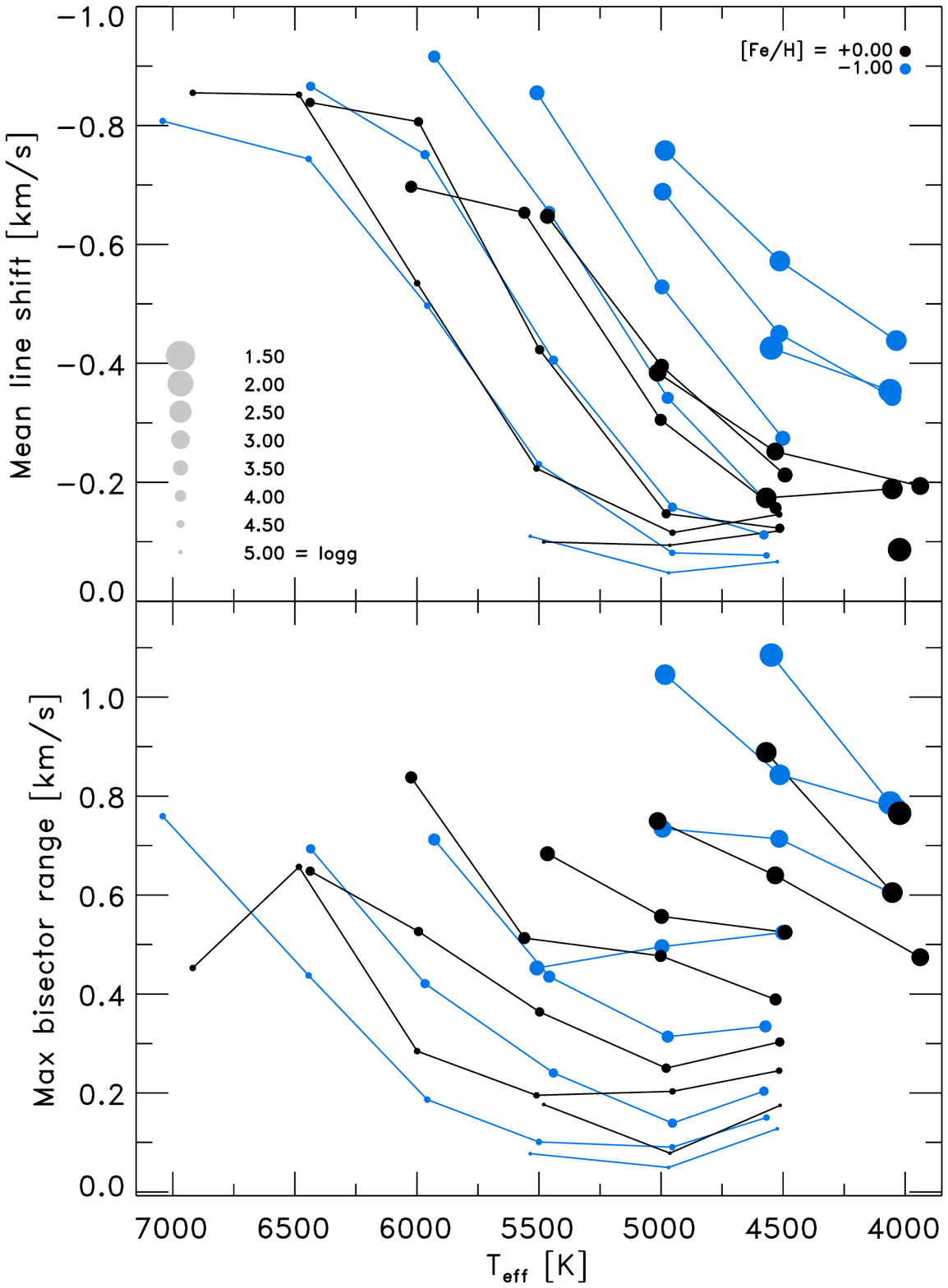}

}

\caption{\emph{Left figure}: Line width $l_{\mathrm{w}}$ (top) and depth $l_{\mathrm{d}}$
(bottom panel) derived from the flux profile of the fictitious $\fei$
line ($\eqw=80\,\mathrm{m\angs},\,\xex=4.0\,\mathrm{eV}$) vs. effective
temperature $\teff$ for different stellar parameters. \emph{Right
figure}: The mean ($\fei$) line shift and maximal bisector for different
stellar parameters. Note the inverted ordinate in the top panel, which
increases towards higher blue-shifts (negative velocity).}
\end{figure*}
In order to study the intrinsic variations of shape of line profiles
with stellar parameters, we determine the line width and depth from
fictitious $\fei$ lines with the same line strength ($\eqw=80\,\mathrm{m\angs}$).
We define the line width, $l_{\mathrm{w}}$ as the full width at half
maximum (FWHM) of the line profile, which is a measure of the range
in Doppler shifts induced by the velocity field. From Fig. \ref{fig:width_depth}
(top panel), one can observe that the line width increases for higher
$\teff$ and $\feh$, and lower $\logg$, which correlates with the
variations of the vertical rms-velocity for the atmosphere models
(see Fig. \ref{fig:psg_velocity_temperature}). The line depth is
the relative flux or intensity of the line core, $l_{\mathrm{d}}=1-\min\left[F_{\lambda}/F_{c}\right]$,
and depicts the maximal absorption from the continuum radiation of
a line. The line depth is clearly anti-correlated with the line width
for different stellar parameters (bottom panel in Fig. \ref{fig:width_depth}).
However, the line depth declines faster than the line width rises,
which signifies that the line profile is becoming flatter and broader,
when considering the same line strength. This broadening of the line
profile for different stellar parameters is a consequence of the higher
velocity amplitudes (convection and oscillations) leading to larger
Doppler shifts. The aspect ratio between depth and width, $a_{\mathrm{dw}}=l_{\mathrm{d}}/l_{\mathrm{w}}$,
diminishes very quickly for hotter $\teff$, lower $\logg$ and $\feh$
for $\fei$ lines. In the case of $\feii$ lines, the variation with
$ $$\teff$ is slightly different, namely the aspect ratio increases
towards higher $\teff$ until $5500\,\mathrm{K}$ and drops above,
and the largest $a_{\mathrm{dw}}$ being slightly smaller. In general
the aspect ratio is increasing with $\eqw$, reaching a maximum around
$50\,\mathrm{m\angs}$, and then decreasing, while for higher $\xex$
it is smaller. The changes of the width, depth and their aspect ratio
are for $\hav$ qualitatively similar, however their amplitude is
rather different. This indicates that the flattening of line profile
is partly due to thermal broadening as well.

\section{Line asymmetry\label{sec:Line-asymmetry}}

\begin{figure*}
\includegraphics[width=176mm]{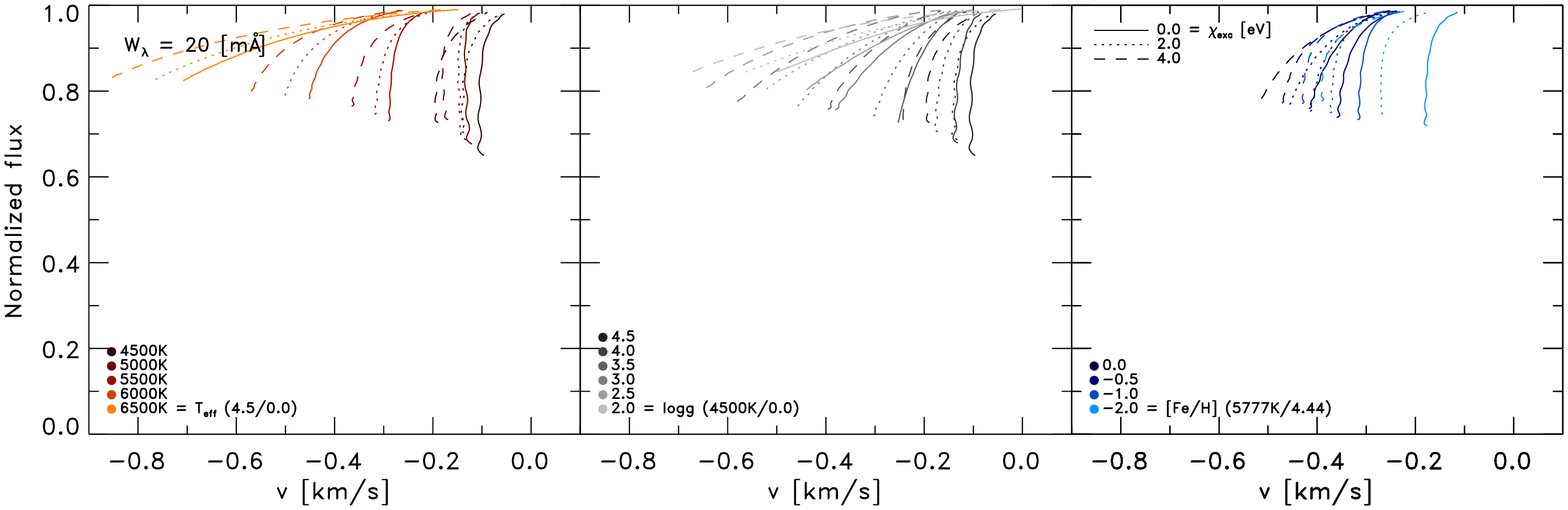}

\includegraphics[width=176mm]{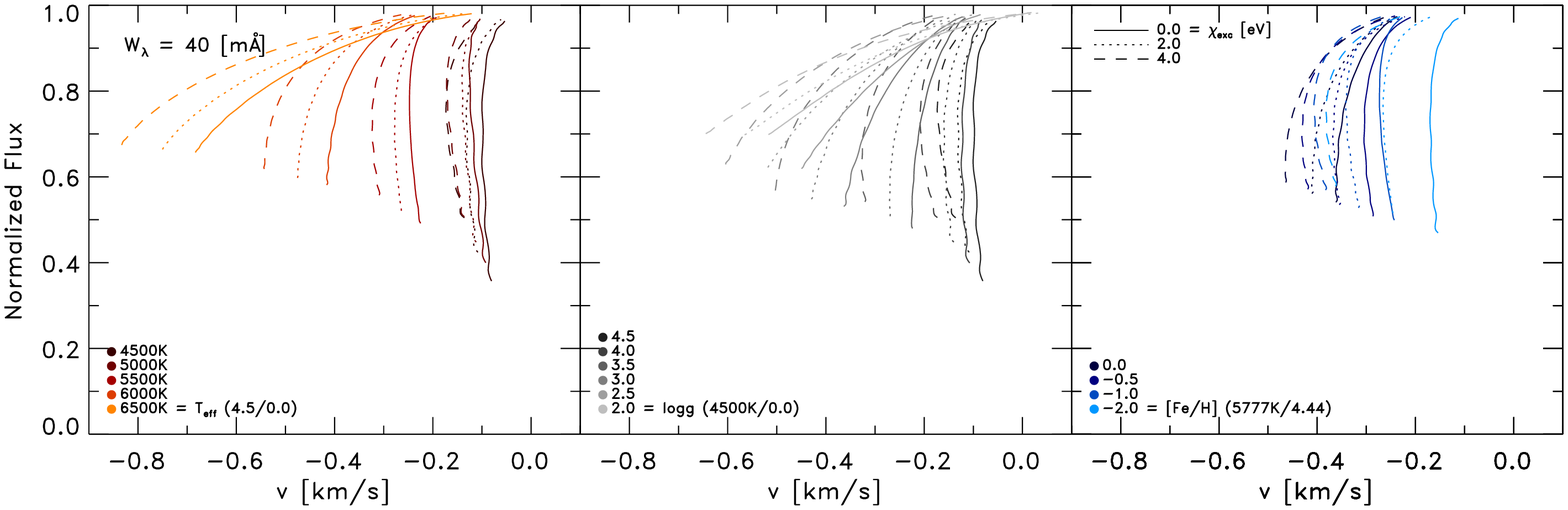}

\includegraphics[width=176mm]{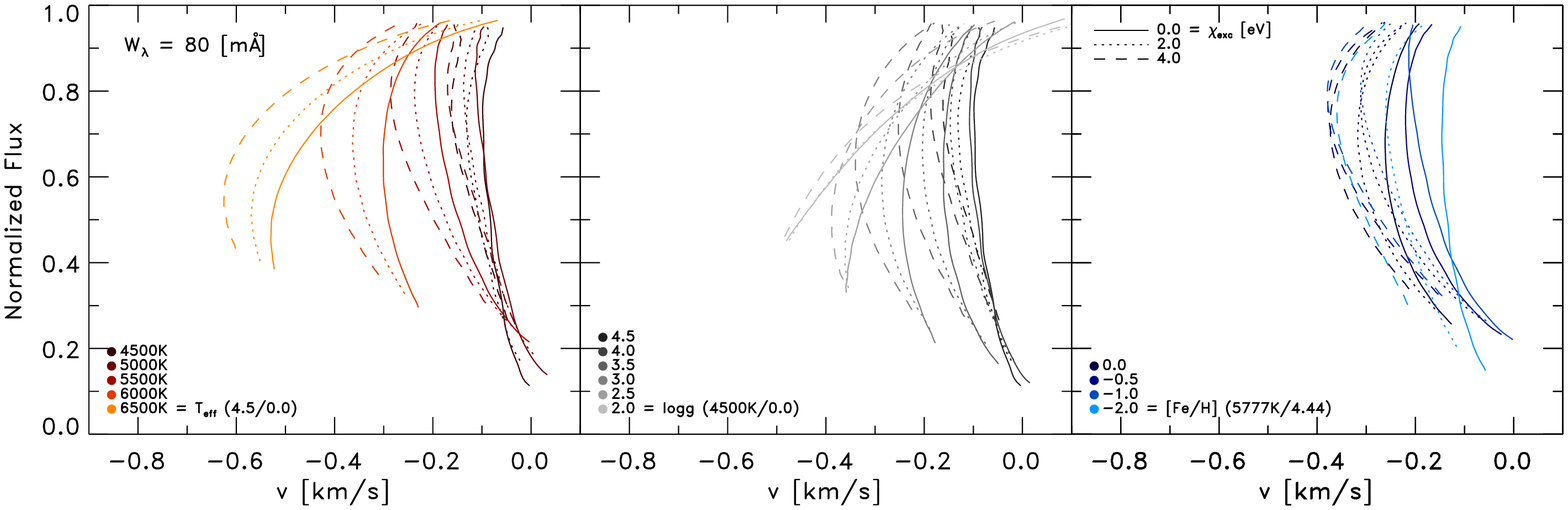}

\caption{Bisectors from fictitious $\fei$ lines with three line strengths
($20,\,40$ and $80\,\mathrm{m\angs}$; top, middle and bottom panel,
respectively) and three different excitation potential energies ($0,2,4\,\mathrm{eV}$;
solid, dotted and dashed lines respectively) for different stellar
parameters. In each column, one stellar parameter is varied, while
the other two are fixed (indicated); \emph{left panel}: effective
temperature; \emph{middle panel}: surface gravity; \emph{right panel}:
metallicity.}

\label{fig:fake_bisector} 
\end{figure*}
The bisector of a spectral line profile is the locus of the midpoints
of the segments identified by the points at constant flux on the line
profile. A line without asymmetries (e.g. 1D or $\hav$ line) has
a straight vertical bisector, while a realistic profile of a spectral
line from a late-type stellar spectrum exhibits a bisector with the
characteristic C-shape \citep{Dravins:1981p15588} that results from
the superposition of the contribution from the strong, blue-shifted,
dominant granules and the weak, red-shifted, intergranular lanes (see
Fig. \ref{fig:line_up_down}).

In Fig. \ref{sec:Line-shape}, we show $\fei$ and $\feii$ bisectors
for different stellar parameters. In general, the typical C-shape
is present in most of the bisectors, however more or less pronounced
depending on the line strength. Weak lines feature a blue-shifted
bisector with a typical $/$-shape depicting only the upper part of
the C-shape, where the line core coincides with the maximal line asymmetry,
$\max\left[\left|v\right|\right]$, i.e. maximal absolute velocity
shift of the entire bisector. On the other hand, the regions of formation
of stronger lines cover larger ranges of optical depths and the cores
of these lines are forming in higher layers above the overshooting
region, therefore, the line centers are less blue-shifted and tend
towards zero, thereby outlining a more defined C-shape. The maximal
asymmetry of strong lines is located around the half height of the
line depth.

For higher $\teff$ the bisectors increase significantly their range,
while the line strength becomes weaker and also more blue-shifted,
until the C-shape finally disappears ($6500\, K$). Towards giants
(lower $\logg$) the line asymmetries (range of bisectors) become
larger in general, and the C-shape is getting more pronounced until
$\logg=3.0$ and less so below that. Furthermore, the upper part of
the bisector recedes towards lower velocity shifts and are even red-shifted
for the lowest surface gravity ($\logg=2.0$), since the contribution
on the red wing from the downdraft is then dominating towards the
continuum flux. With lower metallicity the lines are weaker and the
bisectors lose their C-shape until it vanishes eventually. Also, the
range in bisector diminishes. The variations in the line asymmetry
with stellar parameters result from the differences in line strength,
continuum level, filling factor and Doppler shift. We note that considering
the variations of a single line with different stellar parameters
changes significantly its shape (see Fig. \ref{fig:shape}).

In Fig. \ref{fig:fake_bisector} we illustrate bisectors from the
fictitious $\fei$ line flux profiles for different stellar parameters,
with the same line strength, considering weak ($20,\,40\,\mathrm{m\angs}$)
and intermediate ($80\,\mathrm{m\angs}$) line strength (top, middle
and bottom panel, respectively). The basic idea behind this comparison
is to isolate and illustrate the effect and signature on the line
profile due to the intrinsic structural differences among individual
3D atmosphere models arising solely from the variations in the convective
flow properties. We vary one stellar parameter individually, while
the other two are fixed ($\teff$, $\logg$ and $\feh$). Three different
excitation potential energies ($\xex=0,2,4$~eV) are also considered
(solid, dotted and dashed, respectively).

The intermediate strong lines feature often a more explicit C-shape
compared to the weak lines with smaller maximal bisectors, otherwise,
the changes with stellar parameters are qualitatively rather similar.
For hotter $\teff$ and lower $\logg$, their effect on the weak and
intermediate strong lines is rather similar, namely the line depth
is decreasing (for the same line strength) and the line width is rising,
which means that the line shape becomes increasingly flatter and broader
(see also Sec. \ref{sub:Line-witdh-depth}), while the line shift
and maximal bisector is considerably enhanced, primarily due to concomitant
higher velocity and $T$-contrast (see Sec. \ref{sec:Height-of-line-formation}).
On the other hand, at lower metallicity the changes in line shape
are less pronounced, only the exhibited blue-shifts are lower due
to the smaller level in velocity and $T$-contrast. In the case of
the weak lines (top panel) with lower $\teff$ and $\feh$, higher
$\logg$ the bisectors are increasingly uniform over the entire line
depth, and the C-shape is less distinct, since weak lines are arising
from a smaller extent in height. The variations of the fictitious
$\feii$ are rather similar to those by $\fei$, therefore we refrain
from showing them. The only noteworthy differences are the slightly
smaller ranges in line shift, depth and maximal bisector for fictitious
$\feii$ lines, and the fact that giants feature a stronger influence
from the red-shifted downflows. Lines with higher excitation potential
energy show in general more blue-shifted bisectors, since these lines
form in deeper layers with higher velocity and temperature contrasts.

\begin{figure}
\includegraphics[width=88mm]{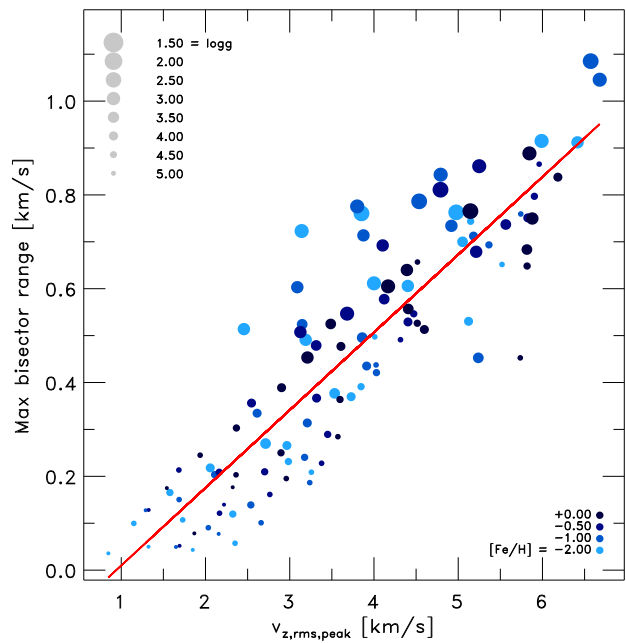}

\caption{The maximal bisector vs. maximal vertical rms-velocity for different
stellar parameters. A linear fit is included (red line).}

\label{fig:max_bisector_velocity} 
\end{figure}
In Fig. \ref{fig:mean_lineshift_max_bisector} (bottom panel) we show
the maximal range of the bisectors with stellar parameters, which
are increasing for higher $\teff$ and $\feh$, lower $\logg$ similar
to the vertical rms-velocity in the 3D atmosphere models (see Fig.
\ref{fig:correlation_velocity_temperature}). One would assume the
line asymmetries correlate with the velocity field, since these arise
from the Doppler shifts. Therefore, we compare the maximal range of
the bisectors with the maximal vertical rms-velocity for different
stellar parameters in Fig. \ref{fig:max_bisector_velocity}. The line
asymmetries correlate clearly with amplitude of the vertical velocity,
only there is a slight scatter due to the different heights of line
formation.

\section{Line shift\label{sec:Line-shift}}

\begin{figure*}
\includegraphics[width=176mm]{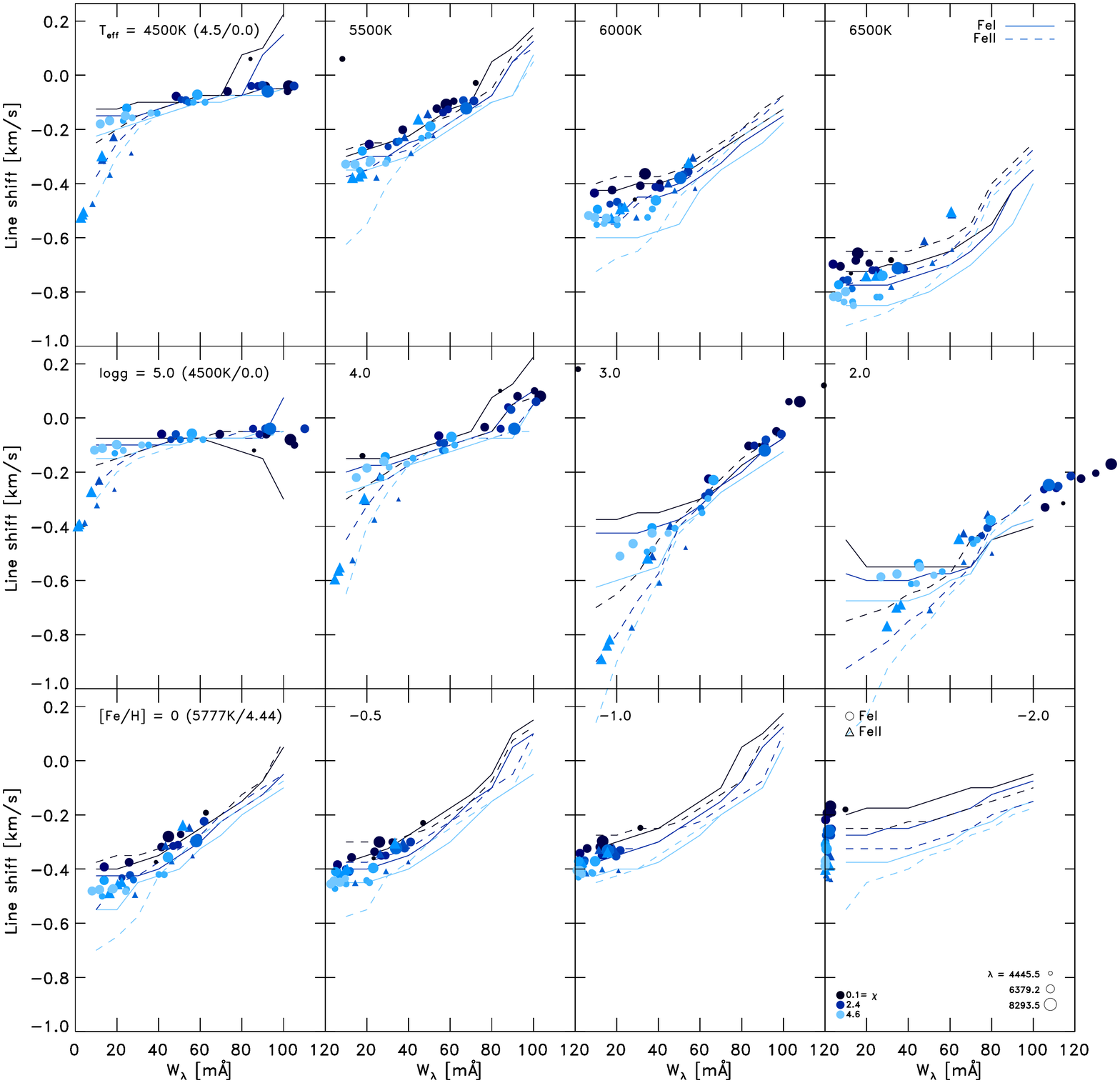}

\caption{Line core shift vs. line strength for the $\fei$ and $\feii$ lines
(circles and triangles, respectively) for different stellar parameter.
Also, the corresponding trends for fictitious $\fei$ and $\feii$
lines are shown (blue solid and dashed lines). The excitation potential
energy $\xex$ (blue colors, where lighter refers to higher $\xex$)
and wavelength $\lambda$ are indicated (symbol sizes, where bigger
refers to higher $\lambda$). In each row, one stellar parameter is
varied, while the other two are fixed (indicated); \emph{top panel}:
effective temperature; \emph{middle panel}: surface gravity; \emph{bottom
panel}: metallicity.}

\label{fig:shift} 
\end{figure*}
In Fig. \ref{fig:mean_lineshift_max_bisector} (top panel), we show
an overview of the mean line shift of the $\fei$ lines with stellar
parameters. Furthermore, we show in Fig. \ref{fig:shift} an overview
of the line shift against the line strength for various stellar parameters,
in order to depict the influence of $\teff$, $\logg$ and $\feh$
individually (from top to bottom respectively) for the complete $\fei$
and $\feii$ line set (circles and triangles respectively). The lower
excitation potential energy, $\xex$, and line wavelength of the lines
are indicated to illustrate trends with line parameters (blue colors
and symbol size respectively).

In the following we discuss the $\fei$ lines first. Towards hotter
effective temperatures (top panel of Fig. \ref{fig:shift}) we find
the maximum shifts of $\fei$ lines to rise considerably (from $\sim0.1$
to $0.8\,\mathrm{km/s}$), while the maximal line strength is diminishing
(from $140$ to $40\,\mathrm{m\angs}$). At higher $\teff$, one finds
convection to operate less efficiently and therefore with more vigorous
flows, with both higher $T$-contrast and rms-velocity (Fig. \ref{fig:correlation_velocity_temperature}).
On the other hand, at higher $\teff$, iron is more likely to be ionized
and also the (continuum) $\mathrm{H}^{-}$-opacity increases, hence,
both effects are reducing the line strength. We note that weaker lines
typically originate from lower depth, where the velocity and $T$-contrast
are larger, therefore imprinting a larger line shift. The line cores
of stronger lines form higher up in the atmosphere, therefore, their
line shifts are smaller (notice the generally smaller line shifts
towards stronger lines in Fig. \ref{fig:shift}). Moreover, for higher
$\teff$ the range in line shift is decreasing. For lower surface
gravity (middle panel), the line shift, its range and the slope of
the linear fits are increasing, reach a maximum at $\logg=3.0$ and
decrease again, while the range in line strength is almost unaffected.
For lower metallicity (bottom panel), the line shifts and line strength
are reducing, and the slope of the linear fit are becoming steeper.

We find in general that the line shifts and strengths are anti-correlated,
i.e. for weaker (stronger) lines their shifts are higher (lower),
which arises from the deeper (higher) location of line formation,
and the respectively larger (lower) velocity amplitude. On the other
hand, lines with lower (higher) excitation potential energy, $\xex$,
exhibit smaller (larger) line shifts. The lines are on average blue-shifted
(negative line shift; see top panel of Fig. \ref{fig:mean_lineshift_max_bisector}),
since the granules occupy a higher filling factors and intensity contribution
over downdrafts. The mean line shift is increasing for higher $\teff$,
lower $\logg$ and enhanced $\feh$. Only a few of the strongest lines
in giants exhibit red-shifted line cores, since the relative contribution
of the red-shifted downflows in these are pronounced. We find that
the fictitious $\fei$ and $\feii$ lines have qualitatively very
similar line shifts as the mean line shifts shown in Fig. \ref{fig:mean_lineshift_max_bisector}.
The latter show a distinct dependence on the line strength.

The $\feii$ lines exhibit in general similar trends for line strength
and shift with stellar parameter as found for the $\fei$ lines. However,
the ranges in line strength are distinctively smaller and its variations
are much less pronounced, in particular the line strength is less
sensitive to $\teff$, since it is the majority species. Furthermore,
the mean line shift and the slope of the linear fits are in general
higher compared to the $\fei$ values.

\section{Conditions at the height of line formation\label{sec:Height-of-line-formation}}

\begin{figure*}
\subfloat[\label{fig:correlation_velocity_temperature}]{\includegraphics[width=88mm]{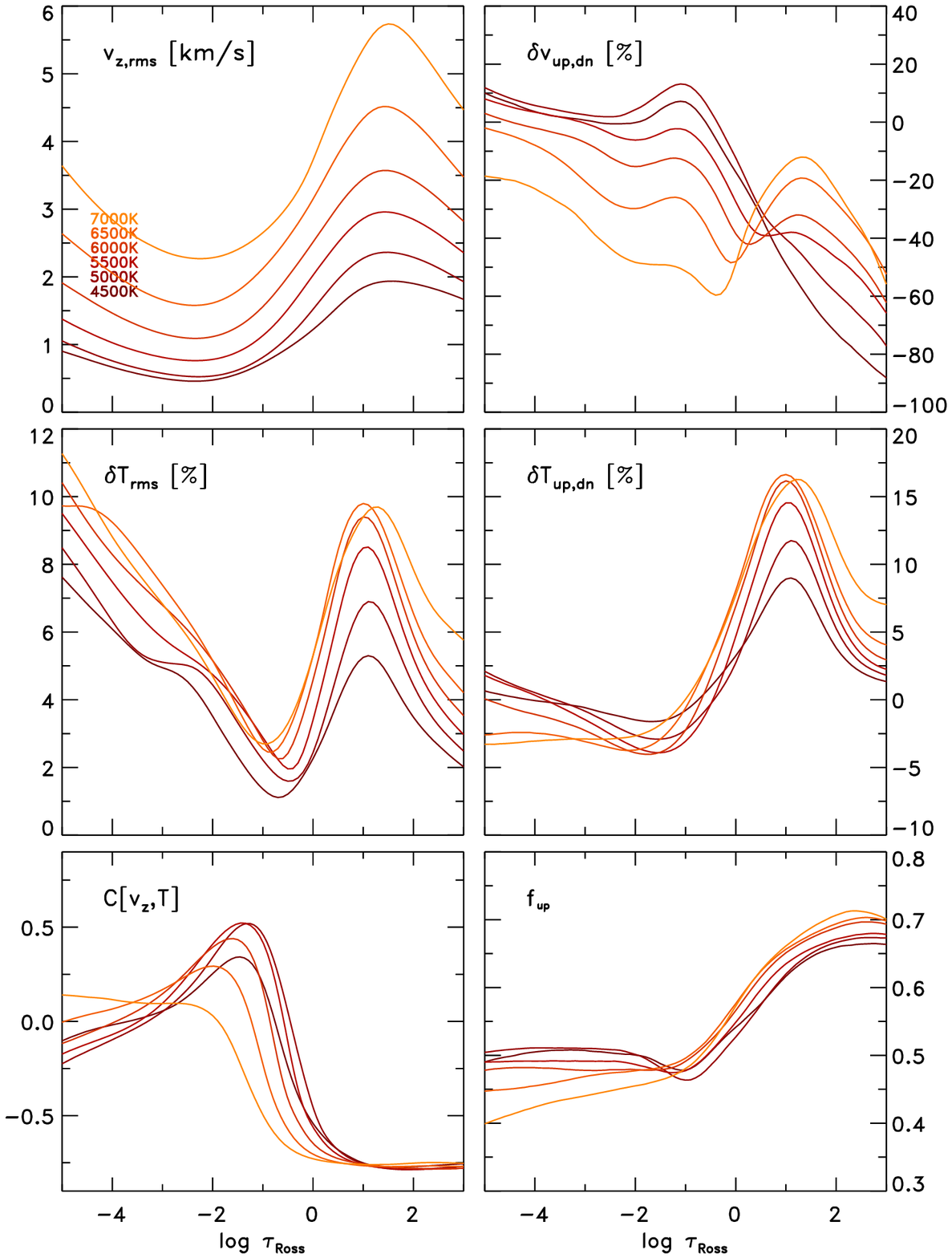}

}\subfloat[\label{fig:psg_velocity_temperature}]{\includegraphics[width=88mm]{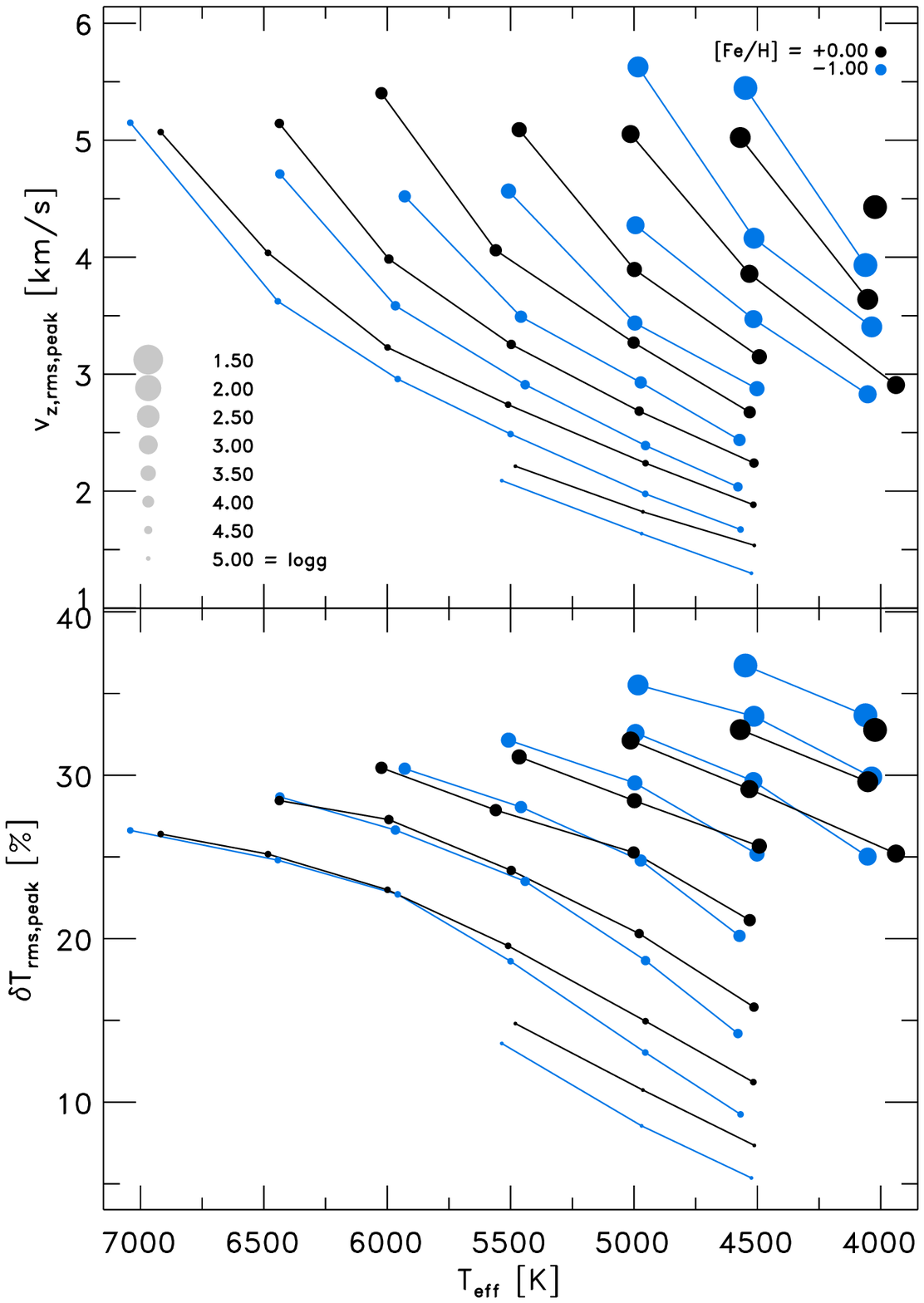}

}

\caption{\emph{Left figure}: The vertical rms-velocity and its asymmetry (top);
$T$-constrast and its asymmetry (middle); correlation function $C\left[v_{z},T\right]$
and filling factor (bottom panel) shown against the optical depth
for models with $\logg=4.5$ and solar metallicity. The different
$\teff$'s are color-coded (red/orange colors). \emph{Right figure}:
Maximal vertical rms-velocity and maximal temperature contrast vs.
$\teff$ (top and bottom panel respectively) for different stellar
parameters.}
\end{figure*}
In the following we want to discuss the different physical conditions
based on the properties of the velocity and temperature prevailing
in the 3D RHD model atmospheres, which are in the end responsible
for the various line asymmetries we seen above. In Fig. \ref{fig:correlation_velocity_temperature},
we show the vertical rms-velocity, $\vzrms$, the asymmetry between
up- and downflow rms-velocity, $\delta v_{\mathrm{up,dn}}$, temperature
contrast, $\delta T_{\mathrm{rms}}$, the temperature difference between
up- and downflows, $\delta T_{\mathrm{up,dn}}$, the correlation function
of the temperature and vertical velocity, $C\left[v_{z},T\right]$,
and the filling factor of the upflows $f_{\mathrm{up}}$.

In the superadiabatic region (SAR) just below the optical surface
($\ltaur>0$) the vertical rms-velocities and temperature contrasts
are reaching their maxima, since at the thin photospheric transition
region the radiative losses from the upflows generate large entropy
fluctuations, that drive stellar surface convection \citep[see paper I and ][ for more details]{Nordlund:2009p4109}.
Further above the top of the convection zone, $\vzrms$ and $\delta T_{\mathrm{rms}}$
decline with height. Also, the asymmetries between the up- and downflows
in velocity and temperature drop fast with height, while in convection
zone, in particular the SAR, $\delta v_{\mathrm{up,dn}}$ and $\delta T_{\mathrm{up,dn}}$
are rather large. Furthermore, below the optical surface, one finds
a tight anti-correlation between the vertical rms-velocity and temperature
due to convective transport of energy, while above the correlation
is distinctively smaller (see $C\left[v_{z},T\right]$ in Fig. \ref{fig:correlation_velocity_temperature}).
The extent of overshooting can be determined with the zero crossing
of the correlation function, $C\left[v_{z},T\right]$, and for higher
$\teff$'s with concomitant higher velocity the overshooting is clearly
shifting towards higher layers. On the other hand, the convective
properties obviously change with stellar parameters. In order to illustrate
this, we show the maxima of $\vzrms$ and $\delta T_{\mathrm{rms}}$
in Fig. \ref{fig:psg_velocity_temperature} for different stellar
parameters. Both quantities clearly increase for hotter $\teff$,
lower $\logg$, and higher metallicity (see Paper I for a detailed
discussion).

Therefore, one finds in general that lines forming in deeper layers
carry larger signatures from the velocity field and temperature contrast,
i.e. resulting in larger line broadening and Doppler shifts, and in
higher layers the opposite is the case. And also the variations of
the line shifts and asymmetries with stellar parameters, which we
discussed above, are in agreement with the properties in the 3D RHD
models.

\section{Conclusions and summary\label{sec:Conclusions}}

We have explored the properties of synthetic spectral lines from neutral
and singly ionized iron in late-type stars with the aid of 3D hydrodynamical
model atmospheres. We have studied the variations with stellar parameters
of aspects such as the strength, width, and depth of spectral lines,
as well as line asymmetries and wavelength shifts. We have related
such variations and the morphology of the asymmetries to the structural
and thermal properties of the 3D models, with particular focus on
velocity and temperature inhomogeneities and their correlation with
depth in the stellar atmosphere.

In Table \ref{tab:line_results} we list our results (line strength,
shift, width, depth and bisectors). A possible application of the
theoretical predictions of the line asymmetries can be the derivation
of radial velocity and gravitational red-shift from high resolution
observations by comparison with 3D line bisectors.
\begin{acknowledgements}
We acknowledge access to computing facilities at the Rechenzentrum
Garching (RZG) of the Max Planck Society and at the Australian National
Computational Infrastructure (NCI) where the simulations were carried
out. Remo Collet is the recipient of an Australian Research Council
Discovery Early Career Researcher Award (project number DE120102940).
\end{acknowledgements}
\bibliographystyle{aa}
\bibliography{papers}

\appendix

\section{Tables\label{app:Tables}}

\begin{table}
\caption{\label{tab:line_list}The he $\fei$ and $\feii$ line parameters
with reference number, ionization degree, wavelength $\lambda$, lower
excitation potential $\xex$, oscillator strength $\lgf$, weighting
factor $g_{u}$, radiation damping $\log\gamma_{\mathrm{rad}}$, lower
level $l_{l}$, upper level $l_{u}$.}

\begin{tabular}{ccccccccc}\hline\hline
 \# & id & $\lambda$ & $\xex$ & $\lgf$ & $g_u$ & $\log\gamma_{\mathrm{rad}}$ & $l_l$ & $l_u$ \\
  1 &   1 &  4445.4717 &      0.087 &     -5.412 &   1 &       4.22 &   s &   p \\
  2 &   1 &  5247.0503 &      0.087 &     -4.961 &   1 &       3.63 &   s &   p \\
  3 &   1 &  5491.8315 &      4.186 &     -2.188 &   2 &       8.09 &   d &   p \\
  4 &   1 &  5600.2242 &      4.260 &     -1.420 &   2 &       8.01 &   p &   s \\
  5 &   1 &  5661.3457 &      4.284 &     -1.756 &   2 &       8.00 &   p &   s \\
  6 &   1 &  5696.0896 &      4.548 &     -1.720 &   2 &       8.33 &   p &   d \\
  7 &   1 &  5705.4648 &      4.301 &     -1.355 &   2 &       8.38 &   p &   s \\
  8 &   1 &  5778.4531 &      2.588 &     -3.440 &   2 &       8.21 &   s &   p \\
  9 &   1 &  5784.6582 &      3.396 &     -2.532 &   3 &       8.05 &   p &   s \\
 10 &   1 &  5855.0767 &      4.608 &     -1.478 &   2 &       8.33 &   p &   d \\
 11 &   1 &  5956.6943 &      0.859 &     -4.552 &   1 &       4.00 &   s &   p \\
 12 &   1 &  6151.6182 &      2.176 &     -3.282 &   1 &       8.29 &   s &   p \\
 13 &   1 &  6240.6460 &      2.223 &     -3.287 &   3 &       6.81 &   s &   p \\
 14 &   1 &  6311.5003 &      2.831 &     -3.141 &   2 &       8.20 &   s &   p \\
 15 &   1 &  6498.9390 &      0.958 &     -4.695 &   1 &       4.36 &   s &   p \\
 16 &   1 &  6518.3671 &      2.831 &     -2.448 &   2 &       8.21 &   s &   p \\
 17 &   1 &  6574.2285 &      0.990 &     -5.010 &   1 &       4.22 &   s &   p \\
 18 &   1 &  6609.1104 &      2.559 &     -2.682 &   1 &       7.99 &   s &   p \\
 19 &   1 &  6699.1416 &      4.593 &     -2.101 &   2 &       8.09 &   s &   p \\
 20 &   1 &  6739.5220 &      1.557 &     -4.794 &   3 &       7.24 &   s &   p \\
 21 &   1 &  6793.2593 &      4.076 &     -2.326 &   2 &       7.56 &   d &   p \\
 22 &   1 &  6837.0059 &      4.593 &     -1.687 &   2 &       7.85 &   s &   p \\
 23 &   1 &  6854.8228 &      4.593 &     -1.926 &   2 &       7.81 &   s &   p \\
 24 &   1 &  7401.6851 &      4.186 &     -1.500 &   2 &       8.01 &   d &   p \\
 25 &   1 &  7912.8670 &      0.859 &     -4.848 &   1 &       3.68 &   s &   p \\
 26 &   1 &  8293.5146 &      3.301 &     -2.203 &   2 &       8.20 &   s &   p \\
 27 &   2 &  4620.5129 &      2.828 &     -3.210 &  31 &       8.56 &   s &   p \\
 28 &   2 &  5264.8042 &      3.230 &     -3.130 &  31 &       8.56 &   s &   p \\
 29 &   2 &  5414.0717 &      3.221 &     -3.580 &  31 &       8.56 &   s &   p \\
 30 &   2 &  6432.6757 &      2.891 &     -3.570 &  31 &       8.49 &   s &   p \\
 31 &   2 &  6516.0767 &      2.891 &     -3.310 &  31 &       8.49 &   s &   p \\
 32 &   2 &  7222.3923 &      3.889 &     -3.260 &  31 &       8.56 &   s &   p \\
 33 &   2 &  7224.4790 &      3.889 &     -3.200 &  31 &       8.56 &   s &   p \\
 34 &   2 &  7515.8309 &      3.903 &     -3.390 &  31 &       8.56 &   s &   p \\
 35 &   2 &  7711.7205 &      3.903 &     -2.500 &  31 &       8.56 &   s &   p \\
\hline\end{tabular}
\end{table}
In Table \ref{tab:line_list}, we present the $\fei$ and $\feii$
line parameters that are used for the line formation calculations,
in the present work. While in Table \ref{tab:line_results} we show
a subset from the main results we presented in our work retrieved
for the solar simulation. The complete list is online available on
CDS. 
\begin{table}
\caption{\label{tab:line_results}Table with main results from synthetic spectral
flux profiles: line strength, width, depth, shift, minimum and maximum
of bisector for the solar simulation. The line number in the first
column is the same as used in Table \ref{tab:line_list}.}

\begin{tabular}{ccccccccc}\hline\hline
 \# & $\eqw $ & $l_{\mathrm{s}}$ & $l_{\mathrm{w}}$ & $l_{\mathrm{d}}$ & min & max \\
  1 &     43.948 &     -0.374 &      4.850 &      0.561 &     -0.373 &     -0.248 \\
  2 &     67.517 &     -0.192 &      5.381 &      0.664 &     -0.276 &     -0.195 \\
  3 &     13.965 &     -0.501 &      4.867 &      0.145 &     -0.500 &     -0.288 \\
  4 &     41.099 &     -0.420 &      5.213 &      0.382 &     -0.432 &     -0.241 \\
  5 &     25.989 &     -0.475 &      5.040 &      0.246 &     -0.473 &     -0.247 \\
  6 &     18.203 &     -0.476 &      5.037 &      0.172 &     -0.490 &     -0.254 \\
  7 &     43.520 &     -0.420 &      5.312 &      0.385 &     -0.421 &     -0.242 \\
  8 &     24.787 &     -0.432 &      4.757 &      0.248 &     -0.431 &     -0.254 \\
  9 &     30.014 &     -0.440 &      4.965 &      0.282 &     -0.434 &     -0.237 \\
 10 &     25.239 &     -0.483 &      5.108 &      0.227 &     -0.473 &     -0.249 \\
 11 &     54.501 &     -0.272 &      4.997 &      0.505 &     -0.303 &     -0.202 \\
 12 &     51.801 &     -0.310 &      5.085 &      0.453 &     -0.332 &     -0.203 \\
 13 &     49.442 &     -0.315 &      5.027 &      0.432 &     -0.337 &     -0.204 \\
 14 &     28.465 &     -0.422 &      4.802 &      0.258 &     -0.412 &     -0.235 \\
 15 &     45.180 &     -0.318 &      4.802 &      0.399 &     -0.321 &     -0.214 \\
 16 &     59.973 &     -0.286 &      5.283 &      0.476 &     -0.308 &     -0.191 \\
 17 &     29.505 &     -0.375 &      4.595 &      0.267 &     -0.371 &     -0.224 \\
 18 &     63.179 &     -0.224 &      5.325 &      0.487 &     -0.295 &     -0.190 \\
 19 &      8.560 &     -0.481 &      4.802 &      0.073 &     -0.481 &     -0.237 \\
 20 &     16.198 &     -0.392 &      4.541 &      0.145 &     -0.403 &     -0.236 \\
 21 &     14.832 &     -0.442 &      4.880 &      0.122 &     -0.443 &     -0.243 \\
 22 &     18.876 &     -0.473 &      4.880 &      0.155 &     -0.463 &     -0.226 \\
 23 &     12.255 &     -0.476 &      4.807 &      0.102 &     -0.474 &     -0.229 \\
 24 &     44.531 &     -0.357 &      5.134 &      0.317 &     -0.360 &     -0.203 \\
 25 &     48.607 &     -0.280 &      4.773 &      0.353 &     -0.291 &     -0.187 \\
 26 &     58.154 &     -0.297 &      5.216 &      0.360 &     -0.306 &     -0.193 \\
 27 &     56.562 &     -0.352 &      5.702 &      0.584 &     -0.425 &     -0.279 \\
 28 &     45.612 &     -0.372 &      5.453 &      0.433 &     -0.414 &     -0.259 \\
 29 &     28.478 &     -0.495 &      5.210 &      0.278 &     -0.487 &     -0.264 \\
 30 &     42.298 &     -0.314 &      5.268 &      0.341 &     -0.344 &     -0.209 \\
 31 &     53.208 &     -0.247 &      5.458 &      0.410 &     -0.306 &     -0.205 \\
 32 &     19.331 &     -0.456 &      5.141 &      0.143 &     -0.470 &     -0.226 \\
 33 &     21.188 &     -0.448 &      5.141 &      0.156 &     -0.457 &     -0.226 \\
 34 &     15.578 &     -0.488 &      5.105 &      0.112 &     -0.483 &     -0.224 \\
 35 &     48.518 &     -0.238 &      5.437 &      0.312 &     -0.304 &     -0.204 \\
\hline\end{tabular}
\end{table}

\end{document}